\documentstyle[12pt,aps,epsf]{revtex}

\def\be{\begin{equation}}
\def\ee{\end{equation}}

\newcounter{fig}

\catcode`\@=11   

\newcommand{\fcaption}[1]{\vspace{1ex}   
        \refstepcounter{figure}   
        \setbox\@tempboxa = \hbox{\footnotesize {\bf Fig.~\thefigure.} #1}   
        \ifdim \wd\@tempboxa > 8cm   
           {\begin{center}   
        \parbox{8cm}{\footnotesize\baselineskip=8pt {\bf Fig.~\thefigure.} #1}   
            \end{center}}   
        \else   
             {\begin{center}   
             {\footnotesize {\bf Fig.~\thefigure.} #1}   
              \end{center}}   
        \fi}

\begin{document}

\title{ \Large \bf Influence of auto-organization and fluctuation effects on the kinetics of a
monomer-monomer catalytic scheme.
}

\bigskip

\author{\Large P.Argyrakis$^{1}$,   S.F.Burlatsky$^{2}$,
E.Cl\'ement$^{3}$, and G.Oshanin$^{4}$}

\address{$^1$ Department of Physics, 
University of Thessaloniki, 54006 Thessaloniki,
 Greece}

\address{$^2$ LSR Technologies, Inc., 898 Main St, Acton, MA 01720-5808
 USA}

\address{$^3$ Laboratoire AOMC, Universit\'e Paris VI, 
4 Place Jussieu, 75252 Paris Cedex 05, France}

\address{$^4$ Laboratoire de Physique Th\'eorique des Liquides, Universit\'e Pierre et
Marie Curie, 4 place Jussieu, 75252 Paris Cedex 05, France}

\bigskip

\address{\rm }
\address{\mbox{ }}
\address{\parbox{16cm}{\rm \mbox{ }\mbox{ }
We study analytically kinetics of an elementary bimolecular reaction scheme
of the
Langmuir-Hinshelwood type
taking place on a $d$-dimensional catalytic substrate.  We propose a general approach which takes into account explicitly the
influence of spatial correlations on the time evolution of particles mean densities
and allows for the analytical analysis. In terms of this approach we recover some of known
results concerning the time evolution of particles mean densities and establish several new ones.
}}
\address{\mbox{ }}
\address{\parbox{16cm}{\rm 
 }}
\maketitle

\makeatletter
\global\@specialpagefalse

\makeatother

\setcounter{page}{1}

\section{Introduction}

Catalytically activated processes play a significant role
 in numerous technologies as they
serve to produce required products from the species which 
are non-reactive 
in normal
physical conditions; these chemically stable species may,
 however, 
 enter into reaction in the presence of some third
substance - the catalytic substrate \cite{eis,zan,avn}. 
Despite of the widespread use of such processes, the
knowledge of the underlying physics and chemistry still
 rests largely on
phenomenological ideas and prescriptions, and thus 
remains a challenge for fundamental
research \cite{eis,zan,avn,osa,good}.  

At the simplest, mean-field level of description,
 reaction kinetics theory presumes 
that the reaction rate should  be
considered as the product of the reactant densities
 and the
rate constant, which is dependent on the nature of 
the binding forces and 
on the particles' dynamics. This rate constant
is proportional to the hopping rate if the process 
is diffusion controlled, or
to the reaction frequency, in case  when the
 reaction is kinetically controlled \cite{smol,col,rice,ovch}. 
Hence, a clear cut
separation is presumed to exist between the local variables, 
 that could be
derived, say, from quantum mechanics, and collective variables, 
expressed in
the most simple way as the product of mean densities of the
 particles involved. 

Recently, one of the
most active developments in the field has been the recognition
 of the substantial
importance of the multi-particle effects,  spatial fluctuations and
 self-organization,
as opposed to conventional local energetic considerations, which emphasized the purely
chemical constraints and focused on the refined descriptions of the elementary reaction
act.
Subsequently, statistical physics
concepts were introduced in order to describe
 anomalous fluctuation-induced behaviors
of non-catalytic chemical reactions
\cite{o,bura,zela,zelb,tous,reda,ov,ova,kop,zhang,avr,cleb,osha,blu,oshd,on} and
simplest catalytic schemes \cite{ziff,fich,cle,red,fla,ficha,clea,mal,klaf,alb,cas}, as
well as such to gain a better understanding of
such collective phenomena as
wave formation, presence of excitable media  or stochastic aggregation in
chemical systems
(see Refs.\cite{cle,clea,mal} and references therein).
One conclusion that can be drawn from the statistical physics approach 
is that fluctuations,
either spatial or temporal, may drive the 
reactive system into a set of new states,
that cannot be understood and described 
in terms of  mean-field kinetic
equations.

In this paper we discuss in detail the influence of spatial fluctuations,
statistical self-organization effects and effects of random diffusive motion of
reactants on the  kinetics of catalytic reactions,
using as a particular example  the  Langmuir-Hinshelwood type 
reaction scheme. 
This reaction process, which is also 
often  referred to as the monomer-monomer
catalytic scheme,  
involves two different kinds of species,
$A$ and $B$, which are deposited (continuously in time with mean
 intensity $I$)
onto the catalytic substrate by an external source, diffuse on the substrate
and react at encounters forming an inert  reaction product
$O$, $A + B \; \rightarrow O$, which is then removed from the system.
Our aims here are twofold. 
First,  we show that this seemingly simple
catalytic reaction  (which has, in fact, several practical
 applications (see, e.g. Ref.\cite{ziff,fich}))
 shows quite a rich behavior
 and represents an ideal
illustrative example of the statistical physics effects 
in the reaction kinetics, which 
may be generic to
more complicated schemes
involved in real "stuff" catalytic processes.
Second, we develop a unified analytical description 
which takes into account explicitly the influence of 
spatial correlations, dimensionality of space
and  the way how the particles are introduced into the reaction bath 
on the time evolution of
observables - mean particle densities. Moreover, our approach  can be routinely generalized 
for kinetic description of more complex reaction schemes.

We focus here on several different aspects of the
Langmuir-Hinshelwood reaction scheme. In particular, we address the question of 
how the kinetics  depends on the
dimensionality $d$ of the reactive system. In addition to the
standard Langmuir-Hinshelwood model in which the catalytic
substrate is a two dimensional flat surface, we analyse 
kinetics of $A + B \rightarrow 0$
reactions followed by an external input of reactive species  in one- and
three-dimensional systems, i.e.  
the situations  appropriate to
reactions in capillary geometries \cite{kopl} and annealing of the radiation damage in
solids \cite{tous}. We show that the monomer-monomer catalytic reaction proceeds
 quite differently  in low dimensional ($d = 1,2$) and
three-dimensional systems. Besides, we examine how the  way
of  particles'
injection into the system affects the properties of  stationary states and also 
how these stationary states are approached in time.
 We consider here two different types of
external input; in the first one (case I) the $A$ and $B$ particles are introduced
independently of each other at random
moments of time and at random positions in space, while in the second case (case II) an $A$ and a $B$
species are introduced in correlated
pairs of a fixed radius, the pairs   being injected
at random moments of time and at random positions in space. We show that also the way
of input does matter significantly and may result in a completely different behavior.

The paper is structured as follows:  In Section II we formulate the model, introduce a general
analytic approach and, in terms of this approach, derive closed-form equations describing
the time evolution of mean particles' densities and pairwise correlation functions. In Section III we present solution
to these equations in one-, two- and three-dimensional systems 
corresponding to different ways of particles injection to the reaction bath.
Finally, we
conclude in Section IV with a brief summary of our results and discussion.

\section{Definitions and basic equations.}

In this section we present kinetic description 
of the monomer-monomer catalytic scheme
 involving diffusive
particles in terms of a certain analytical approach, which takes
 explicitly into
 account the influence of pairwise correlations
on the time evolution of mean densities; such a description 
has been  first proposed in Ref.\cite{bura}, which analysed the effects of
fluctuations on the kinetics of  $A + B \rightarrow O$ reactions and yielded 
the celebrated $t^{-d/4}$-law for the decrease of particles mean
densities. 
The subsequent works
\cite{osha,oshd,on,oshb,bure,burc,burd,mor,mor0,mor1,oshi,oshu} 
extended this approach to more general reaction schemes (e.g. reversible and
 coagulation reactions), three-body and catalytic reactions, 
and also included the possibility of mutual long-ranged interparticle interactions.
Here we will focus on the application 
of this approach
to the analytical description of the  $A + B \rightarrow O$ reactions kinetics 
 in systems with a continuous, random external input of the reactive species. 
We will recover some of already known results obtained for the monomer-monomer
catalytic scheme involving diffusive monomers and will establish
several new ones concerning  the long-time relaxation of mean particles 
density to
their steady-state values and 
the dependence of this steady-state
densities on the system parameters.

We start with the  formulation of the model.
Consider a $d$-dimensional reaction bath of volume $V$
 (we suppose that $V$ is
sufficiently large such that we can discard 
different finite-size effects, e.g. hard-core exclusion between particles,
the termination of
reaction within a finite time interval or saturation (poisoning)) 
in which $A$ and $B$ particles
are continiously introduced by an external random source. The statistical 
properties of
the source will be defined below. After injection to the system, 
the $A$ and $B$ particles begin to diffuse. 
For simplicitly we assume that their diffusion constants are equal to each
other, i.e. $D_{A} = D_{B} = D$. It will be made clear below that such a 
description is also
appropriate to the case of non-equal diffusivities; the calculations in this 
case will be
only essentially more lengthy \cite{oosh}. 
Now, the reaction event is defined as follows:
When any two $A$ and $B$ particles approach each other at a fixed
 separation $R$ 
(the reaction
radius), they may enter into reaction forming (an inert with respect to the reaction)
 reaction product
$O$. The recombination upon an encounter of $A$ and $B$ happens with a finite
probability $p$ (with probability $q = 1- p$ the particles can be reflected)
 which
defines the constant of an elementary reaction act,  $K$. This constant describes the
intrinsic chemical activities of $A$ and $B$  molecules and is dependent on the nature of the
intra- and intermolecular binding forces. In the following we will suppose that this
 purely
"chemical" constant is known $a$ $priori$. Besides, we will assume that  $K$  is
 large (i.e. the probability of particles reflection from each other is low) 
and thus will emphasize the
"statistical physics" effects on the reaction kinetics, 
rather than the effects of chemical  constraints. Consequently, all the factors which
are exponentially small with $K$ will be neglected here.

Now, we define the statistical properties of the particles generation more precisely. 
Let 
$I_{A}(r,t)$ and $I_{B}(r,t)$ be the local, at the point with the radius-vector $r$,
intensities of the production rates of $A$ and $B$ particles. We take that the 
volume average values of the production rates
obey
\be
\frac{1}{V} \; \int_{V} dr 
\; I_{A}(r,t) \; = \; \frac{1}{V} \; \int_{V} dr \; I_{B}(r,t) \; = \; I,
\ee
which means that particles generation is steady in time and the mean 
production rates $I$ of $A$ and $B$ particles are
equal to each other. 

Defining the correlations in the production rates $I_{A,B}(r,t)$ 
we will consider here 
two different situations. In the
first one (case I), we suppose that $A$ and $B$ particles are introduced into 
the reaction bath statistically
independent of each other \cite{ov,ova,kop,zhang,avr,cleb,osha}; that is, 
the fluctuations of the sources are correlated neither in space,
 nor in time. 
In this case, we have
\begin{eqnarray}
\label{2}
&&\frac{1}{V} \; \int_{V} dr \; I_{A}(r,t) \; I_{A}(r + \lambda,t + \tau) \;
 - \; I^{2} \; = \; I \;
\delta(\lambda) \; \delta(\tau),  \; \; \;    (a) \nonumber\\
&&\frac{1}{V} \; \int_{V} dr \; I_{B}(r,t) \; I_{B}(r + \lambda,t + \tau) \; 
- \; I^{2} \; = \; I \;
\delta(\lambda) \; \delta(\tau),  \; \; \; (b) \nonumber\\
&&\frac{1}{V} \; \int_{V} dr \; I_{B}(r,t) \; I_{A}(r + \lambda,t + \tau)  \;
 = \; 0   \; \; \;  (c) 
\end{eqnarray}
In the second case we suppose that $A$ and $B$ particles are introduced 
as correlated $A-B$ pairs\footnote{The particles  in the pair can, of course,  diffuse
apart after injection.}, separated
by a fixed distance $\lambda_{g}$ \cite{ova,avr,cleb,osha}. 
Such type of external generation may arise in chemical systems in which 
 complex reaction product $O$ is continuously forced to break-up by
 an external radiation 
(say, laser pulses)
into the correlated
pairs of the component molecules. 
Here the radius of pair $\lambda_{g}$ will be mainly determined by the
 difference of energy "attributed"
to $O$ and the energy required to dissociate the reaction product.
Another example in which such pairs are produced is the annealing of radiation
 damage in solids. 
When the solid is
irradiated,  atoms are knocked out of their places in the lattice to become
 interstitials
 and leave behind a vacancy, 
and then
the vacancies and interstitials diffuse and recombine. In case II,  
different $A-B$ pairs are statistically uncorrelated and occur,  with an average
intensity $I$,  at random positions
in the reaction bath. Then, the fluctuations of the sources obey \cite{ova,bure}:
\begin{eqnarray}
\label{3}
&&\frac{1}{V} \; \int_{V} dr \; I_{A}(r,t) \; I_{A}(r + \lambda,t + \tau) \; 
- \; I^{2} \; = \; I \;
\delta(\lambda) \; \delta(\tau),   \; \; \;  (a) \nonumber\\
&&\frac{1}{V} \; \int_{V} dr \; I_{B}(r,t) \; I_{B}(r + \lambda,t + \tau) \; 
- \; I^{2} \; = \; I \;
\delta(\lambda) \; \delta(\tau),  \; \; \;  (b) \nonumber\\
&&\frac{1}{V} \; \int_{V} dr \; I_{B}(r,t) \; I_{A}(r + \lambda,t + \tau)  \; = 
\; \gamma_{d}(\lambda_{g})
\; I \; \delta(|\lambda| - \lambda_{g})   \; \; \;  (c)
\end{eqnarray}
In Eq.(3.c), the parameter $\gamma_{d}(\lambda_{g})$ is the normalization factor
 which
 arises because of
different possible angular orientations of a given  $A-B$ pair in a $d$-dimensional
 continuum;
 the value of
$\gamma_{d}(\lambda_{g})$  depends on the dimensionality of the reaction bath
 and for $d = 1, 2$ and $ 3$ equals
respectively $1$, $(2 \pi \lambda_{g})^{-1}$ and $(4 \pi \lambda_{g}^{2})^{-1}$.

Let $C_{A}(r,t)$ and $C_{B}(r,t)$ denote the local densities of $A$
 and $B$ particles at point with
radius-vector $r$ at time $t$. The time evolution of local densities
 due to the diffusion of species,
their reaction and an external production can be described by the following reaction-diffusion equations
\cite{ov,zhang,avr}:
\begin{eqnarray}
\dot{C}_{A}(r,t) \; = \; &-& \; \gamma_{d}(R) \; K \; \int_{V} dr' \; 
\delta(| r - r'| - R) \; C_{A}(r,t) 
\; C_{B}(r',t) \; + \; \nonumber\\
 \; &+& \;  D \; \Delta_{r} \; C_{A}(r,t) \; + \; I_{A}(r,t), 
\end{eqnarray}
\begin{eqnarray}
\dot{C}_{B}(r,t) \; = \; &-& \; \gamma_{d}(R) \; K \; \int_{V} dr' \; 
\delta(| r - r'| - R) \; C_{B}(r,t) 
\; C_{A}(r',t) \; + \nonumber\\
&+& \; D \; \Delta_{r} \; C_{B}(r,t) \; + \; I_{B}(r,t), 
\end{eqnarray}
where the symbol $\Delta_{r}$ denotes the $d$-dimensional Laplace operator
 acting on the spatial variable
$r$,  and the integration with the delta-function $\delta(| r - r'| - R)$ accounts 
for all possible
orientations of an $A-B$ pair, at which an elementary reaction act can
 take place.

Now, an experimentally accessible property is not, however,  the local 
density, 
but rather its volume averaged
value

\be
C(t) \; = \; \frac{1}{V} \; \int_{V} dr \; C_{A,B}(r,t) 
\ee
To find an equation which governs the time evolution of $C(t)$, let us first
 represent the local
densities in the form

\be
C_{A,B}(r,t) \; = \; C(t) \; + \; \delta C_{A,B}(r,t), 
\ee
where $\delta C_{A,B}(r,t)$ will denote local deviations of particles' densities
 from their mean values. By
definition, 

\be
\frac{1}{V} \; \int_{V} dr \; \delta C_{A,B}(r,t) \; = \; 0. 
\ee
Then, substituting Eq.(7) into Eqs.(4) and (5) and taking the volume average,
 we get the following equation

\be
\dot{C}(t) \; = \; - K \; \Big[C^{2}(t) \; + \; G_{AB}(|\lambda| \;  = \; R, t)\Big] \; 
+ \; I, 
\ee
in which $G_{AB}(\lambda, t)$ stands for the pairwise, central correlation function of the form

\be
G_{AB}(\lambda, t) \; = \; \frac{1}{V} \; \int_{V} \; \int_{V} dr \; dr' 
\; \delta( r - r' - \lambda) \;
\delta C_{A}(r,t) \; \delta C_{B}(r',t), 
\ee
the variable $\lambda$ being a $d$-dimensional correlation parameter.

Therefore, Eq.(9) shows that the time evolution of the mean particle density is 
ostensibly coupled to the evolution of the pairwise
correlations in the reactive system. Neglecting these correlations, i.e. 
setting $G_{AB}(\lambda,t) = 0$, which is equivalent
to the traditional, "mean-field" assumption that the particles' spatial
 distribution is uniform, 
we obtain the customary, text-book "law of mass action". 
Such an approximation predicts a linear in time 
growth of mean density
at relatively short times, i.e. 

\be
 C(t) \; \propto \; I \; t, 
\ee
and, in the large-$t$ limit, an exponentially fast relaxation to 
the equilibrium density $C(t=\infty) = (I/K)^{1/2}$, i.e.

\be
C(t) \; \propto \; (\frac{I}{K})^{1/2}  \; \Big[ \; 1 \; - 
\; \exp( - 2 (I K)^{1/2} t) \; + \; ... \; \Big] 
\ee

The short-time behavior as in Eq.(11) is quite reasonable and 
describes the regime in which the particles are
merely added into the (initially empty) system by the external source 
and the reaction between them is negligible, i.e. the regime in which
particle density remains very small. 
As for the analytical 
prediction  in Eq.(12), one may  question its validity on intuitive grounds. 
First,  in the system under consideration diffusive motion of particle is the only
 mechanism to bring
particles together and let them react.  This motion is essentially $d$-dependent 
and should evidently
entail $d$-dimensional behaviors, whilst 
Eq.(12) is independent of the
dimensionality of the reaction bath.
 Second, it
 shows that with an
increase of the chemical reaction constant $K$ the equilibrium density 
tends to zero, 
which is apparently an
 artificial behavior. 
Below we will show that the actual behavior of $C(t)$ as $t \rightarrow \infty$ 
is very different.
This turns out to depend essentially on the dimensionality of the reactive system
 and also on the
way  how the particles are injected into the system. 
 We show analytically that in the case I, when $A$
 and $B$ 
particles are introduced
into the system statistically independ of each other, in low dimensional systems
 (i.e. $d = 1,2$) there occurs a strong reaction-induced spatial 
organization of particles.
 Here the absolute value
of the pairwise correlation function $G_{AB}(R,t)$ (which is negative) grows
 in time, which induces an
unlimited (in absence of hard-core exclusion) growth of mean particles' densities $C(t)$. Therefore,
 in the case I in low dimensional
systems the steady state does not exist, in a striking contradiction to 
the prediction of Eq.(12). Similar results have been obtained previously in
Refs.\cite{ov,ova,kop,zhang,avr,cleb,osha,bure}.
In
the case I, in three dimensional systems the steady state density is 
well-defined but, however,
turns out to be different from that predicted by Eq.(12).  In particular,
 it is dependent on particles'
diffusion constant, which reflects the pathological behavior of correlations 
in the reactive system \cite{ov,ova,osha,bure}.
Actually, we show that correlations of fluctuations in particles local densities
 are essentially
long-ranged and obey, in the steady-state,
 an $\em algebraic$ law $G_{AB}(|\lambda|,t) \sim 1/|\lambda|$.
 Finally, we proceed
to show that in three dimensional systems with uncorrelated generation of particles 
the long-time
relaxation of mean particles density to its steady-state value is described by 
a
 power-law with characteristic exponent $-1/2$, in
contrast to the exponential dependence predicted by Eq.(12). 
Such a behavior has been
first conjectured in Ref.\cite{burd} and here will be deduced analytically.
For the case II,  we show
 that steady-state
density exists in all dimensions \cite{ova,osha,bure}. 
However, its value is also different from that
 predicted by Eq.(12) and
depends, for instance, on the radius of the generated pairs, $\lambda_{g}$. The 
approach to the steady state
in this case is also not exponential in time and is 
 described by a universal algebraic law, reminiscent of the long-time approach to the
equilibrium in reversible chemical reactions
\cite{zela,zelb,oshb,bure,burc,burd,mor}.

We turn back to Eq.(9) and continue our analysis  of the binary reaction kinetics
 taking into account the
influence of pairwise correlations on the time evolution of $C(t)$.
 Omitting the details
 of the derivation,
which can be found in Refs.\cite{bure,burc}, we write down the system of 
equations for the 
time evolution of the pairwise
correlation functions
\begin{eqnarray}
\dot{G}_{AB}(\lambda,t) \; = \; &-& \; K \; C(t) \; \Big[2 \; G_{AB}(\lambda,t) \; + 
 \; G_{AA}(\lambda,t) 
\; + \; G_{BB}(\lambda,t)\Big] \; +  \nonumber\\
&+& \; 2 \; D \; \Delta_{\lambda} \; G_{AB}(\lambda,t) \; + \; I_{AB}(\lambda)
\; + \; T_{AB}(\lambda),
\end{eqnarray}
\begin{eqnarray}
\dot{G}_{AA}(\lambda,t) \; = \; &-& \; 2 \; 
 K \; C(t) \; \Big[ G_{AB}(\lambda,t) \; + 
 \; G_{AA}(\lambda,t)\Big] \; +  \nonumber\\
&+& \; 2 \; D \; \Delta_{\lambda} \; G_{AA}(\lambda,t) \;
 + \; I \; \delta(\lambda)
\; + \; T_{AA}(\lambda), 
\end{eqnarray}
\begin{eqnarray}
\dot{G}_{BB}(\lambda,t) \; = \; &-& \; 2 \;  K \; C(t) \; \Big[ G_{AB}(\lambda,t) \; + 
 \; G_{BB}(\lambda,t)\Big] \; +  \nonumber\\
&+& \; 2 \; D \; \Delta_{\lambda} \; G_{BB}(\lambda,t) \;
 +  
 \; I \; \delta(\lambda)
\; + \; T_{BB}(\lambda), 
\end{eqnarray}
where $\Delta_{\lambda}$ denotes the  Laplace operator acting on
 the $d$-dimensional variable $\lambda$,
the symbol $I_{AB}(\lambda)$ in Eq.(13) describes the correlations in the production
 rates of $A$ and $B$
particles; it is equal to zero in the case I (uncorrelated generation of particles) 
and to

\be
I_{AB}(\lambda) \; = \; \gamma_{d}(\lambda_{g}) \; I \; \delta(|\lambda| - 
\lambda_{g})  
\ee
in case II, when particles are introduced into the system in correlated $A-B$ pairs.
 Finally, in Eqs.(13) to
(15) the terms $T_{ij}$ denote the correlation functions of the third order. 

The time evolution of the pairwise correlations is coupled to the evolution
 of the third-order correlations, which, in turn, depends on the correlations of
 the fourth order. Thus the non-linearity of the reaction-diffusion Eqs.(4) and (5)
entails an infininite hierarchy of equations for correlation
functions and one has to resort to some approximate methods. 

The most commonly 
used method invokes truncation of this hierarchy approximating the third-order 
correlation
 functions in terms of $C(t)$ and $G_{ij}(\lambda,t)$ \cite{kuz}. Such an approach,
 as it
 was first noticed in Ref.\cite{tous}, results in the Smoluchowski-type approximate results
 with
 improved numerical coefficients and is appropriate for the
 description of the single-species reactions $A + A \rightarrow O$, but not for the
 description of reactions involving two different types of particles. The point is that
 such an approximation misses an important conservation law, which is specific for 
$A + B \rightarrow O$ reactions. Namely, the reaction process conserves the difference
 $Z(r,t)$ of local densities, $Z(r,t) = C_{A}(r,t) - C_{B}(r,t)$, which changes
 only
 due to the diffusion of particles and thus is a pure diffusive mode of the system. 
Conservation of $Z(r,t)$ entails, in turn, the conservation of the combination of
 pairwise correlation functions, $S_{-}(\lambda,t) = G_{AA}(\lambda,t) + 
G_{BB}(\lambda,t) - 2 G_{AB}(\lambda,t)$, which is also a pure diffusive 
mode\footnote{One may readily verify \cite{bure} 
that $T_{AA} + T_{BB} - 2 T_{AB}$ is exactly
 equal to zero.}.  Consequently, only such truncation scheme should correctly describe 
the behavior of the binary reaction, which does not violate the important conservation laws  \cite{tous}.

The most simple truncation scheme, which preserves the conservation laws, has
 been  first 
proposed in Ref.\cite{bura}. In this scheme the third-order correlation functions, 
i.e. $T_{ij}$, were set equal to zero. Such a truncation, as it was shown in 
Refs.\cite{osha} and \cite{oshb,bure,burc} is equivalent to the assumption that fields 
$\delta C_{A,B}(\lambda,t)$ have Gaussian distribution. Then, the fourth-order 
correlation functions automatically decouple into the product of pairwise correlation
 functions and the third-order correlations are equal to zero. Such an approach leads 
to, for instance, the correct long-time decay law of the particles mean densities,
 i.e. the $t^{-d/4}$-law, 
but fails to reproduce correctly the intermediate time behaviors; at intermediate times
this approach predicts essentially the same behavior as the formal kinetic
 "law of mass
action" and thus disregards the effects of particles' diffusion at the intermediate-time stage. 

This shortcomming has been revisited and improved in Refs.\cite{oshb,bure,burc},
 in which it
has been shown that correlation functions of the third order are small only 
in the
limit $|\lambda| > R$, while in the domain  $|\lambda| \approx R$ they  are 
singular
and this singularity has an impact on the behavior of the pairwise correlation
 at the
intermediate times. In a discrete-space picture, essential at scales 
$|\lambda| \approx R$, the third-order correlation functions have been computed
explicitly \cite{bure},

\be
T_{AA} \; \approx \; T_{BB} \; \approx \; T_{AB} \; \approx \; \gamma_{d}(R) \;
\dot{C}(t) \; \delta(|\lambda| - R).   
\ee

Substituting Eq.(17) into Eqs.(13) to (15) one gets then a closed with respect to $C(t)$
and $G_{ij}(\lambda,t)$ system of equations. To solve them, it is expedient to represent
the pairwise correlations in the form

\be
G_{ij}(\lambda,t) \; = \; \hat{G}_{ij}(\lambda,t) \; + \; g_{ij}(\lambda,t), 
\ee
where $\hat{G}_{ij}(\lambda,t)$ denotes a "singular" part of the pairwise correlation
functions, which accounts merely for the behavior of the third-order correlations, and  
$g_{ij}(\lambda,t)$ - the "fluctuational" part, which accounts for the
fluctuation spectrum of the external source and fluctuations stemming out of reaction and
diffusive processes. 

The "singular" part of the pairwise correlation functions has been determined in
Ref.\cite{bure}. In particular, the leading at sufficiently large
times behavior of $\hat{G}_{ij}(|\lambda| = R,t)$ have been found to be as follows:

\be
\hat{G}_{ij}(R,t) \; \approx \; \dot{C}(t) \; (\pi t/ 8 D)^{1/2}, 
\ee
in one-dimensional, 
\be
\hat{G}_{ij}(R,t) \; \approx \; \dot{C}(t) \; \frac{ln(D t/R^{2})}{8 \pi D}, 
\ee
and
\be
\hat{G}_{ij}(R,t) \; \approx \; \dot{C}(t) \; 8 \pi D R, 
\ee
in two-, and three-dimensional systems, respectively.

Now, inserting Eqs.(19) to (21) to Eq.(9) we obtain the following equation
 for the time
evolution of particles mean density

\be
\dot{C}(t) \; = \; - \; \frac{K  K_{S}(d)}{K + K_{S}(d)} \; \Big[ C^{2}(t) \; + \;
g_{AB}(|\lambda| = R,t)\Big] \; + \; \frac{I}{1 + K/K_{S}(d)}  
\ee
where $K_{S}(d)$ obeys, as $t \to \infty$,

\be
K_{S}(d = 1) \; \approx \; (\frac{8 D}{\pi t})^{1/2},
\ee
\be
K_{S}(d = 2) \; \approx \; \frac{8 \pi D}{ln(D t/R^{2})}, 
\ee
and
\be
K_{S}(d = 3) \; \approx \; 8 \pi D R
\ee

One may readily notice that in three-dimensions the $K_{S}(d)$, Eq.(25),
 coincides exactly with
the so-called "diffusive" Smoluchowski 
constant; a reaction constant which has been first calculated by
Smoluchowski \cite{smol}
in his approximate description
 of the effects of diffusion on the chemical reactions kinetics. This
constant accounts for, heuristically, the "resistivity"
 of random, diffusive transport of particles with
respect to reaction. Employing the Smoluchowski method, the analogues of 
such a constant have been obtained in Ref.\cite{tor}
for one- and two-dimensional systems. 
Remarkably, our results in  Eqs.(23) and (24) coincide with those obtained in
Refs.\cite{ovch} and \cite{tor}. We note also that the prefactor before the brackets in Eq.(22), i.e.
the ratio $K_{app} = K  K_{S}(d)/( K + K_{S}(d))$, is the so-called effective
 or apparent
reaction constant, which was  first derived for three-dimensional systems in
Ref.\cite{col}. Therefore, an account of the "singular" part of the third-order 
correlation
function and subsequent extraction of the "singular" part in the pairwise correlators
leads us to the results equivalent to those obtained in 
terms of the Smoluchowski
approach. 

Hence, Eq.(12), in which one sets $g_{AB}(|\lambda| = R,t)= 0$ and $K_{S}(d)
=\infty$ reduces to the formal kinetic "law of mass action", 
while setting    
$g_{AB}(|\lambda| = R,t)= 0$ and using $K_{S}(d)$ as in Eqs.(23) to (25), one
obtains
the effective kinetic equation of
 the Smoluchowski-type approach. Below, we proceed to show
that taking into account the time evolution of the pairwise correlations, i.e. 
the term $g_{AB}(|\lambda| = R,t)$, one arrives at completely different
physical behavior as compared 
to the ones predicted by the formal kinetic and
Smoluchowski approaches.  

Finally, we obtain the following  system of
equations, obeyed by the "fluctuational" part of the pairwise correlation functions.
 It
reads
\begin{eqnarray}
\dot{g}_{AB}(\lambda,t) \; = \; &-& \; K \; C(t) \; \Big[2 \; g_{AB}(\lambda,t) \; + 
 \; g_{AA}(\lambda,t) 
\; + \; g_{BB}(\lambda,t)\Big] \; +  \nonumber\\
&+& \; 2 \; D \; \Delta_{\lambda} \; g_{AB}(\lambda,t) \; + \; I_{AB}(\lambda),
\end{eqnarray}
\begin{eqnarray}
\dot{g}_{AA}(\lambda,t) \; = \; - \; 2 \;  K \; C(t) \; \Big[ g_{AB}(\lambda,t) \; 
+ \; g_{AA}(\lambda,t)\Big] \; + \; 2 \; D \; \Delta_{\lambda} \; g_{AA}(\lambda,t) \; 
+ 
 \; I \; \delta(\lambda), 
\end{eqnarray}
\begin{eqnarray}
\dot{g}_{BB}(\lambda,t) \; = \; - \; 2 \;  K \; C(t) \; \Big[ g_{AB}(\lambda,t) \; + 
 \; g_{BB}(\lambda,t)\Big] \; + \; 2 \; D \; \Delta_{\lambda} \; g_{BB}(\lambda,t) \; 
+  
 \; I \; \delta(\lambda)
\end{eqnarray}
Equations (26) to (28), accompanied by Eq.(22),  
are now closed with respect to mean particles' densities and pairwise
correlations, and suffice the computation of the time evolution of the monomer-monomer reaction scheme.

\section{Kinetics of the monomer-monomer reaction scheme.}

Below we will analyse solutions of Eqs.(22) to (28) in systems of different
dimensionalities and with different types of external particle generation. 
The derivation of results in case of one-dimensional systems will be presented in
detail. 
The steps involved for such a
derivation in higher dimensions are essentially the same and here we will merely
discuss the results. 

\subsection{Low dimensional systems.}

Let us start with the case of one-dimensional systems in which an external source
produces uncorrelated $A$ and $B$ particles. 

We note first that the system of equations (26) to (28) possesses two integrable
combinations
\be
S_{-}(\lambda,t) \; = \; 2 g_{AB}(\lambda,t) \; - \; g_{AA}(\lambda,t) \; -
\; g_{BB}(\lambda,t),
\ee
which is related to the conserved property $Z(r,t)$, and
\be
S_{+}(\lambda,t) \; = \; 2 g_{AB}(\lambda,t) \; + \; g_{AA}(\lambda,t) \; +
\; g_{BB}(\lambda,t) 
\ee
These integrable combinations  obey
\be
\dot{S}_{-}(\lambda,t) \; = \; 2 D \Delta_{\lambda} S_{-}(\lambda,t) \; - 
\; 2 \; I
\delta(\lambda), 
\ee
which is thus the pure diffusive mode of the system, not affected by the reaction, 
and
\be
\dot{S}_{+}(\lambda,t) \; = \; 2 D \Delta_{\lambda} S_{+}(\lambda,t) \; 
- \; 4 K
C(t) S_{+}(\lambda,t) \; + \; 2 \; I \delta(\lambda) 
\ee
The desired property, i.e. the correlation function $g_{AB}(\lambda,t)$ which
enters the Eq.(22), may be then expressed in terms of these integrable combinations as
\be
g_{AB}(\lambda,t) \; = \; \frac{1}{4} (S_{-}(\lambda,t) \; + \; S_{+}(\lambda,t)) 
\ee

Consider now the solutions to Eqs.(31) and (32) in one-dimensional systems. 
Applying the Fourier transformation
over the variable $\lambda$,
\be
S_{\pm}(\omega,t) \; = \; \frac{1}{\sqrt{2 \pi}} \int^{\infty}_{-\infty} d\lambda \;
\exp(i \omega \lambda) \; S_{\pm}(\lambda,t) 
\ee
to Eqs.(31) and (32), and assuming that at $t = 0$ no $A$ and $B$ particles were
present in the system, 
one readily gets that the Fourier-images of the integrable
combinations follow
\begin{eqnarray}
S_{-}(\omega,t) \; &=& \; - \; I \; \sqrt{\frac{2}{ \pi}} \; \int^{t}_{0} d\tau 
\; exp( - 2
D \tau \omega^{2}) = \nonumber\\
 &=& \; - \; \frac{I}{ D \omega^{2} \sqrt{2 \pi}} \; ( 1 \; - 
\; exp( - 2 D \tau \omega^{2})),   
\end{eqnarray}
and
\be
S_{+}(\omega,t) \; = \; - \; I \; \sqrt{\frac{2}{ \pi}} 
\; \int^{t}_{0} d\tau \; exp( - 2
D \tau \omega^{2} \; - \; 4 K \int^{t}_{\tau} d\tau' \; C(\tau')) 
\ee
Now we notice that in the extreme situation, when reaction act occurs at any encounter
of any $A$ and $B$ particle (i.e. when $K = \infty$) the second integrable combination
$S_{+}(\omega,t)$ vanishes since the integral $\int^{t}_{\tau} d\tau' \; C(\tau')$ is
obviously positively defined. One can show, however, that even for the finite $K$s the
influence of $S_{+}(\omega,t)$ on the pairwise correlation function, Eq.(33), is not
essential at large times and the dominant contribution to $g_{AB}$ comes from 
$S_{-}(\omega,t)$.  

We note that setting $t = \infty$ in Eq.(35) we obtain that $S_{-}(\omega,\infty)$
has a steady-state spectrum of the form $S_{-}(\omega,\infty) \sim 1/\omega^{2}$, i.e.
the spectrum which has an essential singularity when $\omega \rightarrow 0$. Such a 
singular behavior of the fluctuation spectrum of the pairwise correlations in
 systems with
binary reactions followed by an external uncorrelated production of the reactive
species has been first
obtained, in terms of a different than ours approach, in Refs.\cite{ov} and \cite{ova}. The authors
concluded thus that the steady state of such a system is highly anomalous; since such a
singularity is not integrable in low dimensional systems, the steady state values of the
integrable combination $S_{-}(\lambda,t)$ and thus of the correlation function $g_{AB}$
are infinitely large, which means that as time evolves the system progressively
coarses into the domains containing particles of only one type. 

Consider now how the integrable combination $S_{-}(\lambda,t)$ and the correlation
function $g_{AB}$ grow in time. Taking the inverse Fourier transformation of the first
line in Eq.(35) we get
\be
S_{-}(\omega,t) \; = \; - \; \frac{I}{\sqrt{2 \pi D}} \; \int^{t}_{0} \frac{d\tau
}{\sqrt{\tau}}\; \exp(
- \lambda^{2}/8 D \tau) 
\ee

The integrand in Eq.(37) is a bell-shaped function with its maximum at point $\tau =
\lambda^{2}/8 D$. For bounded $\lambda$,  the bulk contribution to the integral 
comes from the
algebraic tail $1/\sqrt{\tau}$ and consequently, the leading at $t \gg R^{2}/8 D$
behavior of the integrable combination follows
\be
S_{-}(|\lambda| = R,t) \; = \; - \; I \; \sqrt{\frac{2 t}{\pi D}} 
\ee
Accordingly, the absolute value of the "fluctuational" part of the pairwise correlation
function grows in time as
\be
g_{AB}(|\lambda| = R,t) \; = \; - \; I \; \sqrt{\frac{t}{8 \pi D}}  
\ee
Inserting the just derived expression into the Eq.(22) and noticing that the correct
asymptotical behavior of the mean density obtains when the terms in brackets 
compensate each other, i.e., when
\be
C(t) \; = \; \sqrt{ - g_{AB}(|\lambda| = R,t)}, 
\ee
we find that
\be
C(t) \; = \;  I^{1/2} \; (\frac{t}{8 \pi D})^{1/4},  
\ee 
i.e. in one-dimensional systems with random uncorrelated generation of the reactive
species the mean particle density grows sublinearly in time as time progresses.
The behavior as in Eq.(41)  has been also obtained
earlier in Refs.\cite{ova,kop,zhang,avr,cleb,osha}. 

Consider now how the situation will be changed in the case II, when $A$ and $B$
particles are introduced into the reactive bath as correlated pairs. In this case one
readily gets that the Fourier-image of the integrable combination $S_{-}(\lambda,t)$
obeys the following equation
\be
\dot{S}_{-}(\omega,t) \; = \; - 2 D \omega^{2} S_{-}(\omega,t) \; - \; 2 \; I \; (1
\; - \; cos(\omega \lambda_{g}))
\ee 
whose solution will read
\be
S_{-}(\omega,t) \; = \; - \;  
\frac{I \; (1 \; - \; cos(\omega \lambda_{g}))}{ D \omega^{2} \sqrt{2 \pi}} \; (1 \; -
\; \exp( - 2 D t \omega^{2}))    
\ee
One may readily notice a very important feature of Eq.(43);
 in a striking contrast to the case I, here the steady-state 
spectrum is not singular
in the limit $\omega \rightarrow 0$, but tends to a constant value
\be
S_{-}(\omega \rightarrow 0,t = \infty) \; = \; - \; \frac{I \;
\lambda_{g}^{2}}{\sqrt{8 \pi} D}, 
\ee
which means that $S_{-}(\lambda,\infty)$ and hence, $g_{AB}(\lambda,\infty)$ are
 bounded in
systems of any dimensionality, and thus the well-defined steady-state mean
density $C(t = \infty)$ also exists. We notice, however, that the steady-state
pairwise correlation function is proportional to $\lambda_{g}^{2}$ and thus may increase
indefinitely with growth of $\lambda_{g}$. This unbounded growth is, of course, quite
consistent with the result in Eq.(39), since the limit $\lambda_{g} \rightarrow \infty$
corresponds to the case of uncorrelated generation of particles. 

Now, the inverse Fourier transformation gives us 
\begin{eqnarray}
S_{-}(\lambda,t) \; &=& \;  - \; \frac{I}{\sqrt{2 \pi D}} \int^{t}_{0}
\frac{d\tau}{\sqrt{\tau}} \; \Big\{ \exp( - \frac{\lambda^{2}}{8 D \tau}) \; - \nonumber\\
 &-& \; \frac{1}{2} \exp(- \frac{(\lambda - \lambda_{g})^{2}}{8 D \tau})
\; - \; \frac{1}{2} \exp(- \frac{(\lambda + \lambda_{g})^{2}}{8 D \tau})\Big\} \; = \nonumber\\
 &=& \; - \; \frac{I \lambda}{4 D \sqrt{\pi}} \;
\Big\{\Gamma(-1/2,\frac{\lambda^{2}}{8 D t}) \; - 
\; \frac{1}{2} \Gamma(-1/2,\frac{(\lambda_{g} - \lambda)^{2}}{8 D t}) 
\; - \nonumber\\
&-& \; \frac{1}{2} \Gamma(-1/2,\frac{(\lambda_{g} + \lambda)^{2}}{8 D t})\Big\}, 
\end{eqnarray}
where $\Gamma(\alpha,x)$ denotes the incomplete Gamma function \cite{abr}. 

Consider now
the asymptotic behavior of the pairwise correlation function, Eq.(33), for different
values of parameters $\lambda$ and $\lambda_{g}$, and different values of time $t$.

One readily gets from Eq.(45) that at sufficiently short times, when $\lambda \gg
\lambda_{g} \gg 8 D t$, the pairwise correlation function obeys
\be
g_{AB}(\lambda,\lambda_{g},t) \; = \; - \; \frac{3 I \lambda_{g}^{2} (8 D t)^{3/2}}{8
\sqrt{\pi} D \lambda^{4}} \; \exp( - \frac{\lambda^{2}}{8 Dt}), 
\ee
which shows that correlations drop off as a Gaussian function at large scales. 

Now, at
short scales, such that $\lambda \ll \lambda_{g}$ and $\lambda \ll 8 D t$, 
and when $\lambda_{g}$ is
sufficiently large,  $\lambda_{g} \gg 8 D t$, we obtain that
$g_{AB}(\lambda,\lambda_{g},t)$ obeys Eq.(39), which is not a surprising result since
at such scales the correlations in the particles' injection should be irrelevant. 

Within the opposite limit, when $8 D t \ll \lambda \ll \lambda_{g}$ the correlator
follows
\be
g_{AB}(\lambda,\lambda_{g},t) \; \approx \; \; - \; \frac{I  (8 D t)^{3/2}}{16
\sqrt{\pi} D \lambda^{2}} \; \exp( - \frac{\lambda^{2}}{8 Dt}), 
\ee
which is reminiscent of the behavior in Eq.(46). 

Finally, in the limit when both $\lambda \ll 8 D t$ and $\lambda_{g} \ll 8 D t$ (and
$\lambda_{g} > \lambda$), 
i.e. in the limit of very long
times, we find the following asymptotic expansion
\be
g_{AB}(\lambda,\lambda_{g},t) \; 
 \approx  \; - \; \frac{I  (\lambda_{g} -
\lambda)}{16
D} \; \Big[ 1 \; - \; \frac{\lambda_{g}^{2}}{\lambda \sqrt{\pi D t}} 
\exp( - \frac{\lambda^{2}}{8 Dt})\; + 
\; {\mathcal O}(1/t) \Big], \;
\lambda_{g} \geq \lambda,
\ee
where the symbol ${\mathcal O}(1/t)$ signifies that the correction terms decay with time as $1/t$.

Equation (48) suffices to derive the large-$t$ evolution of the mean particle density in
the case of generation by correlated pairs, which reads
\be
C(t)  \; \approx \;  \sqrt{\frac{I  (\lambda_{g} -
R)}{16
D}} \; \Big[ 1 \; - \; \frac{\lambda_{g}^{2}}{2 R \sqrt{\pi D t}} 
 \; + 
\; {\mathcal O}(1/t) \Big], \;
\lambda_{g} \geq R 
\ee

Equations (49) reveals two surprising features; first, the steady-state density
turns out to be dependent both on the diffusion constant and on the radius of pairs,
generated by the source. Such a dependence is, of course, inconsistent with the
predictions of the formal kinetic approach, based on the text-book "law of mass action".
 Second, the approach of particles' densities to
their steady-state values obeys a power-law dependence, in a striking contrast to the
exponential one predicted both by the formal kinetic and the Smoluchowski approach.

To close this subsection let us briefly consider the behavior of solutions of the 
reaction-diffusion equations
(22), (26) to (28) in two-dimensional systems.

In the case I, we have from Eqs.(31) and (33) that as $t \rightarrow \infty$ the
pairwise correlation function grows (by absolute value) as
\be
g_{AB}(|\lambda| = R, t) \; \approx \;  - \; I \; ln(D t/R^{2}), \; D t  \; \gg
\; R^{2}, 
\ee
and consequently, we get from Eq.(40) that in this case at large times
the mean particle density
exhibits logarithmically slow growth \cite{ov,ova,kop,zhang,avr,cleb,osha},
\be
 C(t) \; \approx \; \sqrt{ I \; ln(D t/R^{2}) }
\ee
Now, in the case II, we obtain that the steady-state exists and $g_{AB}(|\lambda| = R,
\infty)$ behaves as
\be
g_{AB}(|\lambda| = R, \infty) \; \approx \; - \; I \; ln(\lambda_{g}), 
\ee
and hence, the steady-state density turns to be a slowly growing function of the
radius of the generated pairs,
\be
C(t = \infty) \; \approx \; \sqrt{ I \; ln(\lambda_{g})}  
\ee
Some analysis shows also that such a steady-state is approached via an algebraic law
\be
C(t) \; - \; C(t = \infty) \; \approx \; (D t)^{-1}, 
\ee
in contrast to the exponential in time dependence predicted by mean-field descriptions.

\subsection{Three-dimensional systems.}

As we have already mentioned, in the case I the steady-state fluctuation spectrum is
characterized by an essential singularity of the type $1/\omega^{2}$ as $\omega
\rightarrow 0$. In three-dimensional systems such a singularity is integrable, which
insures that the steady-state correlations exist and vanish as
$\lambda \rightarrow \infty$.  Consequently, the steady-state mean particle
density $C(t = \infty)$ should exist in 3D also in this case. Let us analyse now
the form of these correlations. Solving Eq.(31) in 3D we find (up to the correction
terms which are exponentially small with 
$K$) that
\be
g_{AB}(\lambda, t = \infty) \; \approx \; - \; \frac{I}{8 \pi D \lambda}, 
\ee
i.e. $A-B$ correlations vanish with the distance between particles $\lambda$ as
$1/\lambda$, which shows that in the monomer-monomer catalytic scheme taking place in
three dimensional systems the
correlations in the steady-state show a quasi-long-range order decaying  as the first
inverse power of the interparticle distance.

Now, substituting Eq.(55) into the Eq.(22) we find the following expression for the
steady-state density in 3D,
\be
C(t = \infty) \; = \; \sqrt{(\frac{1}{8 \pi D R} \; + \; \frac{1}{ K}) \; I}, 
\ee
which shows that $A-B$ correlations lead here to an effective renormalization of the reaction constant
in the steady-state, i.e. $C(t = \infty)$ has the form $C(t = \infty) =
\sqrt{I/K_{app}}$, where $K_{app}$ is the mentioned above
apparent reaction constant \cite{col}. 

Consider now how such a steady-state is approached at long times. 
Expanding the solution of Eq.(31)
near the steady-state, we have that pairwise correlations approach the steady-state as
a power-law, 
\be
g_{AB}(|\lambda| = R,t) \; \approx \; 
\frac{I }{8 \pi D R } \;  \; \Big[ 1 \; - \; \frac{R}{ \sqrt{\pi D
t}} \; + \; O(1/t)\Big],
\ee 
which yields, in turn, a power-law relaxation of the mean particle density to the
steady-state
\be
C(t) \; - \; C(t = \infty) \; \approx \;  (D t)^{-1/2} 
\ee

Therefore, in contrast to low dimensional systems, in three dimensional systems with
random uncorrelated generation of the reactive species the essential singularity in the
fluctuation spectrum is integrable, correlations vanish with the distance between
particles and the
steady-state mean particle
density exists. However, the steady-state density is different from that predicted by the mean-field "law
of mass action" and shows, in particular, dependence on the particles' diffusivity $D$. Besides,
Eq.(58) reveals that here approach to the steady-state is
described by a power-law with the characteristic exponent $-1/2$, which is essentially
non-mean-field behavior.

Finally, for the case II we find the following results for the correlation function and
mean density. In the steady-state the $A-B$ correlations are equal to zero
 for $\lambda \geq
\lambda_{g}$ (again, apart from the exponentially small with $K$ terms). In the domain
$\lambda < \lambda_{g}$ the correlations exist and are described by
\be
g_{AB}(\lambda,t = \infty) \; \approx \; - \; \frac{I}{8 \pi D \lambda} \; (1 \; - \;
\frac{\lambda}{\lambda_{g}}),
\ee  
which reduces to the result in Eq.(55) when $\lambda_{g} = \infty$. 
In constrast to the
behavior as in Eq.(55), however, the correlations vanish at finite values of the
correlation parameter $\lambda$. 

Now, Eq.(59) yields for
the steady-state mean-particle density 
\be
C(t = \infty) \; = \; \sqrt{I \; (\frac{1}{K} \; + \; \frac{1 \; - \;
R/\lambda_{g}}{8 \pi D R})}, 
\ee
which is less than the steady-state density in the case I, Eq.(56), due to a factor $1 -
R/\lambda_{g}$, which renormalizes the Smoluchowski constant. 
Consequently, for $\lambda_g > R$ apparent rate constant here takes the form
\be
K_{app} = \frac{8 \pi D R K}{8 \pi D R + (1 - R/\lambda_g) K}
\ee
We find also that such a steady-state is approached via a power-law,
\be
C(t) \; - \; C(t = \infty) \; \approx \; (D t)^{-3/2}  
\ee
which is faster than the approach described in Eq.(58), but still very different from the exponential behavior predicted
by mean-field analysis.

\section{Conclusion.}

To summarize, we have shown that both  the cases I and II 
the fluctuations effects dominate the
kinetics of the monomer-monomer catalytic scheme involving diffusive particles and
induce essential departures from the predictions of the mean-field approaches. In the
case I, the effects of fluctuation are especially pronounced in low dimensional systems -
the steady-state does not exist and mean particle density grow indefinitely in
time, in absence of hard-core exclusion between particles.
 In three dimensions the steady-state exists, but is characterized by very strong
interparticle correlations, which, in turn, have a strong impact on the value of the
steady-state mean particle density. The steady-state density is different from that predicted by mean-field "law
of mass action". 
The approach to this steady-state is
described by an anomalous power-law with the characteristic exponent $-1/2$, which
stems from the presence of essential singularity in the steady-state fluctuation spectrum. In
the case II, the steady-state fluctuation spectrum and the steady-state mean particle
density exist in any dimension, but show an anomalous, non-mean-fields dependence
on the particles' diffusivity and the radius of pairs, generated by the source.
Approach to the steady-state follows a universal power-law with the characteristic
exponent $- d/2$, which resembles, apart from the dependence of the prefactors on
the system parameters (e.g. constant of the backward reaction)
the long-time approach to the equilibrium in
reversible chemical reactions \cite{zela,zelb,oshb,bure,burc,burd,mor}. The origin of
such a behavior is  that the fluctuation spectrum in the steady-state is flat at
small values of the wave-vector, i.e. the essential singularity in the steady-state
spectrum of fluctuations is screened.

\bigskip

{\Large Figure Caption.}

Fig.1. Langmuir-Hinshelwood reaction on a two-dimensional catalytic 
substrate. Black and grey spheres denote particles of 
$A$ and $B$ species, respectively;
$(1)$ describes the situation in which an $A$ and a $B$ appear
  within the
reactive distance from each other and may enter into reaction.

\end{document}